# Starting Life and Searching for Life on Rocky Planets

*Paul B Rimmer[1], Sukrit Ranjan[2], Sarah Rugheimer[3]*

**ABSTRACT**
The study of origins of life on Earth and the search for life on other planets are closely linked. Prebiotic chemical scenarios can help prioritize planets as targets for the search for life as we know it and can provide informative priors to help us assess the likelihood that particular spectroscopic features are evidence of life. The prerequisites for origins scenarios themselves predict spectral signatures. The interplay between origins research and the search for extraterrestrial life must start with lab work guiding exploratory ventures in the solar system, and the discoveries in the solar system informing future exoplanet observations and laboratory research. Subsequent exoplanet research will in turn provide statistical context to conclusions about the nature and origins of life.

1. Department of Earth Sciences, University of Cambridge, Cambridge, UK

2. CIERA, Northwestern, USA

3. Department of Physics, University of Oxford, Oxford, UK

**LIFE BEYOND EARTH: LESSONS FOR THE ORIGINS OF LIFE?**

The search for life beyond Earth offers the tantalizing possibility of constraining theories of the origin of life (abiogenesis). Most generally, the search for life on potentially habitable worlds, whether positive or negative, offers the best opportunity to constrain the abiogenesis rate (i.e., the fraction of habitable worlds which give rise to life; Chen & Kipping 2018). The abiogenesis rate further informs origins theories. If future observations suggest that the abiogenesis rate is high, then it suggests the mechanisms by which life emerges to be simple and high-probability, whereas if it turns out to be low, then complex, low-probability origin-of-life scenarios must be considered.

More specifically, some origins theories require specific planetary environments in order to achieve their proposed outcomes (Benner et al. 2012; Russell et al. 2014; Sasselov et al. 2020). If life is detected on a planet which fails to meet such requirements, then it demonstrates the corresponding theory is not the only pathway to abiogenesis. The search for signals of extraterrestrial life provides the *only* opportunity for tests of abiogenesis theories under conditions which embrace the heterogeneity and complexity of realistic environments over geologic scales and time. Abiogenesis theories here are cast in terms of **prebiotic chemical scenarios** or **origins of life scenarios**, by which we mean a network of chemical reactions and mechanisms (e.g. formation of amino acids by shock chemistry of $CH_4$, $NH_3$ and $H_2O$) that take place in a particular geochemical context (e.g. lightning in a chemically reduced atmosphere above a warm pond).

The solar system features numerous potentially-habitable environments being scrutinized as venues for life (see Schwieterman et al. 2018 and references therein), and a thorough review is beyond the scope of this work. Here, we focus on reviewing the potential of the solar system planets as probes of prebiotic chemistry (chemistry that appears to be relevant to the origin of life). Next to Earth, the most intensely studied planet is Mars.

The focus on Mars has been motivated in part by geological markers of a wet and possibly warm climate in its past, and in part by potential for habitable environments on modern Mars (discussed by Benner at al. 2020). Detection of life on Mars would be an epochal discovery, revolutionizing our understanding of planetary habitability. However, discovery of Martian life would not necessarily constrain our understanding of abiogenesis, due to the possibility of lithopanspermia (i.e., transfer of life in the form of spores between planets by meteorites; Melosh 1988). Mars and Earth interchange meteoritic material, raising the possibility that putative Martian life may be the product of terrestrial contamination (or vice versa; see Benner et al. 2020 for a recent review). The real value of Mars for the study of prebiotic chemistry may come from (1) its lack of hydrologic and tectonic activity, which means that geologic imprints of its early history may be preserved on its surface, and (2) the potential similarity between primitive terrestrial and Martian environments, meaning that studies of Martian geology may constrain terrestrial prebiotic conditions, and hence studies of prebiotic chemistry on Earth (Sasselov et al. 2020; Benner et al. 2020). Putative Venusian life faces similar concerns; however, Venus is a sufficiently hostile environment that it is possible that life there requires a fundamentally different biochemistry (e.g., alternate bioscaffold; Petkowski et al. 2020). If so, detection of Venusian life might be attributable to a second abiogenesis.

In the outer solar system, the risk of transfer of spores from the surface of Earth due to meteorites (lithopanspermic transfer) from Earth is significantly lower than in the inner solar system. Travel times from the terrestrial planets are much longer, and the probability of spore survival concomitantly lower, though it is hard to falsify this proposition entirely (Worth et al. 2013).  As with inner solar system worlds, these caveats could be eliminated if detected life utilizes fundamentally different biochemistry (e.g., utilizing a non-aqueous solvent, utilizing a non-carbon scaffold). We focus on two prebiotically interesting subsets of the class of potentially-habitable outer solar system worlds: ocean worlds and a hydrocarbon world. First, some moons of the gas giant, like Enceladus (moon of Saturn) and Europa (moon of Jupiter), while not possessing atmospheres, host subsurface oceans that are potentially habitable. We are not aware of proposals for habitable surface-like niches present on such worlds over geologically long times. Detection of life on such worlds might therefore test whether abiogenesis is possible in oceanic conditions (Russell et al. 2014). Second, at the cold temperatures present on the surface of Titan, water is a mineral, but lakes of liquid hydrocarbon with active hydrologic cycles abound (Hayes et al. 2016). Titan's reducing atmosphere is conducive to Miller-Urey style prebiotic chemistries (Horst et al. 2012).

It is therefore tempting to speculate as to the possibility of life utilizing non-aqueous solvents on Titan. This speculation is relevant because it can allow for the widening of the habitable zone (the zone within which life on exoplanets may have a good chance of being detected) beyond the liquid water habitable zone (the zone within which liquid water can exist stably on the surface of a rocky planet). Detection of life thriving within the methane lakes of Titan would falsify water as a bio-requirement. In some sense, detection of non-aqueous life would align with the naive expectation from simple chemistry, whereby water might be considered unfavorable as a biotic solvent due to the susceptibility of many organics to hydrolysis (Benner et al. 2012).

It is unlikely that life on exoplanets and on Earth had a common origin: the transfer timescales are too long, the transfer masses too small, and the prospect of a common origin with terrestrial life can be reasonably excluded (Chen & Kipping 2018). Exoplanets also offer the prospect of using large sample sizes to marginalize over ambiguities in biosignature detections, possibly permitting inference regarding the presence of life from a sample of planets even when the data are insufficient to infer strong conclusions from any single planet (Bean et al. 2017). Consequently, the search for life on exoplanets offers the possibility of strong tests of theories of abiogenesis. Such tests must be carefully considered. For example: Habitable rocky planets around stars much cooler than our Sun, such as M-dwarfs, have surface environments with much less ultraviolet light. Therefore, detection of life on planets around these stars might test whether UV is required for the emergence of life (Ranjan et al. 2017; Rimmer et al. 2018). Rimmer et al. (2018) experimentally characterize a low bound on the UV irradiation required for the cynanosulfidic scenario to function and report that no known M-dwarf has adequate UV output to meet this threshold. Ranjan et al. (2017) find instead that higher-mass, more active M-dwarfs may have adequate UV, which would imply a wider abiogenesis zone, more favorable for M dwarfs in general. The difference may relate to the calculation of atmospheric attenuation in these models; efforts to reconcile these calculations are underway. Another important caveat is that M-dwarf flares, through direct or indirect production of UV, might substitute for steady-state stellar UV; further work is needed to explore this possibility, and verify whether M-dwarf planets can in fact test whether UV is required for the origin of life. Similarly, some exoplanets may be "water worlds" which lack dry land (Ramirez & Levi 2018). Future observations from such worlds may test the proposition that dry land is required for the origin of life. Such questions may seem premature, since we have not yet detected life elsewhere. However, we may draw a parallel to gravitational waves, whose potential as physical and cosmological probes was investigated well in advance of their detection. It may be worthwhile to similarly consider the potential of exo-biological detections as tests of basic theories. More generally, theories of abiogenesis are likely to be incorporated into exoplanet life search as priors to help infer whether detection of a potential biosignature gas can best be attributed to biological or abiological causes (Catling et al. 2018). Remote observation of spectroscopic features of prebiotically relevant molecules on young exoplanets may help to further inform these priors.

**PREBIOSIGNATURES ON YOUNG EXOPLANETS**

Prebiosignatures are signs that a planet hosts environment favorable for prebiotic chemistry. There are a variety of surface prebiosignatures relevant for solar system investigation, especially on Mars (Sasselov et al. 2020) and the Moon. We will focus on atmospheric prebiosignatures, molecules that can be present in an exoplanet atmosphere and that can be detected spectroscopically. These molecules are either useful for prebiotic chemistry in themselves (primary prebiosignatures) or they indicate the current or recent occurrence of global physical processes that can be conducive for prebiotic chemistry (secondary prebiosignatures). These physical processes include volcanism, lightning, meteoritic impacts, and high fluxes of stellar energetic particles (see https://doi.org/10.7910/DVN/UTFCHI)

Consider hydrogen cyanide (HCN) as an example. Hydrogen cyanide is both a primary and secondary prebiosignature. It is a primary biosignature because it is a reactant at the beginning of many prebiotic chemical scenarios (e.g., Ruiz-Mirazo et al. 2014). It is also a tracer of impact chemistry, lightning chemistry in reduced atmospheres and chemistry driven by stellar energetic particles (see https://doi.org/10.7910/DVN/UTFCHI). Two parameters worth considering for prebiotic chemistry are the local C/O and H/C ratios. The C/O ratio must be >~ 1 in order for HCN and other prebiotically relevant species to be generated photochemically or by other disequilibrium processes (Rimmer & Rugheimer 2019) and the H/C ratio must be <~ 2 in order to form other prebiotically relevant species, such as $HC_3N$ (see references in https://doi.org/10.7910/DVN/UTFCHI).

Figure 2 shows a list of prebiosignatures relevant for several origins scenarios, indicating the wavelength ranges and relative strengths of their spectroscopic features, whether they are primary or secondary biosignatures, and if they are secondary biosignatures, what physical process or processes they trace.

Distributions of prebiosignatures in planetary atmospheres for young systems are useful for determining what fraction of planets likely had large environmental areas conducive to prebiotic chemistry. This will help inform prior probabilities when considering candidate biosignatures on evolved planets, and will be useful when comparing distributions of prebiosignatures and biosignatures in the future.

**BIOSIGNATURES ON EVOLVED WORLDS**

*Conventional Biosignatures:*

Biosignatures in the exoplanet context are typically gases that are indicators of life and are detectable from light years away. Often biology uses the same chemical energy sources as chemistry and geology, and so biosignatures often require context to distinguish them from signs of life or some non-biological process.

The classic biosignatures first proposed were oxygen or ozone in combination with methane (see Schwieterman et al. 2018 and references therein for a review). It is thought that methanogens, microbes producing $CH_4$, were abundant on early Earth. Upon its advent, oxygenic photosynthesis quickly became the most successful biomass building metabolism on Earth, producing abundant $O_2$. It is the combination of these two gases which is considered a robust biosignature; while abiotic mechanisms can maintain either in isolation in an atmosphere, we have not yet identified abiotic mechanisms which can simultaneously maintain significant atmospheric abundances of both oxygen and methane. To our knowledge, only life is capable of this feat; and hence it is the disequilibrium combination of these two gases that embodies the "gold standard" for atmospheric biosignatures, though a challenging one to measure.

Oxygen and its photochemical byproduct, ozone, can be produced by UV radiation splitting apart $H_2O$ or $CO_2$, and methane can be produced by serpentinization or by volcanoes (see e.g. Schwieterman et al. 2018). However, discriminating whether or not a potential biosignature is from life can be achieved by confirming the presence or absence of other molecules, as well as a robust understanding of the stellar radiation environment (e.g. Rugheimer & Kaltenegger 2018). Detecting signs of life on an exoplanet will require gathering as much context as possible of the planetary environment, such as: is the planet rocky and is its surface temperate? Figure 3 shows some scenarios for detection, false positives and discriminates for oxygen as a biosignature.

One of the important contexts for biosignatures is the presence of liquid water. As discussed above, all life as we know it uses water as its solvent. Water is abundant in the Universe, is stable in liquid form over a wide range of pressures and temperatures, and is a polar molecule, which allows it to be a good solvent. These facts combined motivate using water to define habitability and guide our first steps in searching for life beyond Earth. Detection of life in alternate solvents (e.g., the hydrocarbon lakes of Titan) would permit broadening of this criterion.

Life has been intertwined in the geological evolution of Earth. Earth immediately after formation was molten and heavily bombarded with material from the planetary disk. As it cooled, liquid water became stable on the surface. It is uncertain when life originated, but there is debated isotopic evidence dating back to 4.1 billion years ago and robust evidence of stromatolites at 3.5 billion years (Javaux 2019 and references therein).

The rise of oxygen at 2.33 Ga (Luo et al. 2016) fundamentally changed the planet's biosphere and geology. From an exoplanet biosignature perspective, Earth's living biosphere was most detectable after oxygen rose to where its photochemical byproduct ozone reached detectable levels, and methane was still abundant. The balance between how detectable ozone and methane are in combination depends both on the production of each molecule and the stellar radiation environment and resultant photochemistry (Rugheimer & Kaltenegger 2018).

*Novel Biosignatures:*

In addition to the conventional biosignatures which have been well-studied in extant literature, novel biosignatures in the search for life have been proposed, such as dimethyl sulfide, $CH_3Cl$, $N_2O$, and $NH_3$ in a hydrogen dominated world (see Schwieterman et al. 2018 and references therein). Ideally a biosignature is a gas that is produced by life, has strong spectral features, and has few or no false positives. Seager et al. (2016) have proposed a systematic search of all small molecules for those that fit these criteria. Such work is challenging, because for most of these molecules we do not have an accurate understanding yet of the chemical networks that could create these molecules in other environments. Additionally, the spectra are not known for over 99% of the potential molecules of interest; efforts are underway to meet this challenge (Sousa-Silva et al. 2019). Finally, it is unknown how many of these molecules will be produced globally at sufficiently high rates by ecosystems on other planets (exobiospheres) to build up to sufficient concentrations to be observed by future telescopes at a distance of light-years. The productivity of putative exo-biospheres can ultimately only be constrained by observations: theoretical work to enable these observations remains to be completed.

*Abiosignatures:*

As biosignatures are signs of life, abiosignatures are signs of its absence. Candidate abiosignatures are often species in thermochemical equilibrium that provide a potential source of free energy for life, such as $H_2$ and CO (Sholes et al. 2019). The reasoning is that if there were a well-established biosphere, it would consume all of these species and expel species such as $CO_2$ and $H_2O$ as waste. Their presence indicates an absence of this biosphere, or at least minor activity of the biosphere. It is unclear whether CO in particular qualifies as an abiosignature, given that it is produced by life on Earth and that it is likely to build up on planets around cool stars at levels that would overwhelm biological consumption (Schwieterman et al. 2019). CO is also predicted to have been quite abundant in simulated biospheres on the inhabited Archean Earth (Kharecha et al. 2005).

While prebiosignatures might indicate a planet with habitable conditions, ripe to proceed down an evolution of life path, it is also possible that these molecules would represent a failed biosphere. Prebiosignatues such as hydrogen cyanide provide a significant source of energy for the organisms that had originated via its consumption. Hydrogen cyanide in particular is not known to be a major waste product of metabolism, and so the presence of HCN may provide some evidence of life's absence.

**LIFE AS WE DON'T KNOW IT**

This article has focused on starting and finding life as we know it. What about life as we don't know it? This can mean a variety of different things. It can be about the composition of life: life that is made of different carbon-based building blocks, either slightly different (Carell et al. 2012) or wildly different (Hud & Anet 2000), or life that is made of different elements altogether (Petkowski et al. 2020). It can be about the environment in which life finds itself, which may be very different from any historical environment on Earth[1]. It may also deal with the molecules produced by life (Seager et al. 2016), or about observations of molecular complexity that are agnostic about specific molecular species (Marshall et al. 2017).

If life can be built out of different carbon-based materials, then the transitional prebiosignatures, between the starting materials and life's building blocks, could be different than discussed above. However, many alternatives considered, such as nucleotides with sugars other than ribose or deoxyribose (XNA), alternative nucleobases, or other amino acids, are still built out of largely the same molecules: simple sugars (e.g., from formaldehyde, glycolaldehyde), bases in the form of molecular rings made of nitrogen and carbon atoms (e.g., purine or pyrimidine), that can result from HCN chemistry, and molecular species that resemble amino acids, the building blocks of proteins. All these have the same sorts of starting materials and tracers as listed in Figure 2.

Life has been hypothesized to originate and exist in environments very different from Earth's, from the global and deep ocean of Europa, which is almost completely sealed off from the surface by many kilometers of ice, in the sulfuric acid clouds of Venus, in the frigid methane lakes of Titan, or floating in water droplets suspended in the clouds of Jupiter (see McKay 2020 and references therein). Life that exists in these places, especially where water is not a solvent, will require an alternative biochemistry (Petkowski et al. 2020).

The search for life as we don't know it is impossible within our own solar system, even on Earth, without first exploring the chemistry in the lab. Once the chemical mechanisms involved with hypothetical life as we don't know it are worked out in the lab, in situ exploration can take place on other planets, ideally supported by remote observational evidence, then the search for exotic life on exoplanets can commence with some hope of success. Even if an alternative biochemistry is never discovered, the exploration in these environments is of intrinsic interest. Agnostic biosignature criteria based on a mathematical estimate of chemical complexity have been proposed (Marshall et al. 2017) and these also must be tested, first in the lab, then on Earth, and then searched for on objects in our own solar system, before there can be a successful search for this exotic life on exoplanets. The reason for this is a matter of resolution. Consider hypothetical life in hydrocarbon lakes as an example. Without establishing the high-resolution context accessible first in the lab and then on planets and moons close enough to send missions, we will have no basis for evaluating a spectral signature on an exoplanet with surface hydrocarbon lakes to evaluate whether it is due to (a) abiotic processes, (b) biological processes from aqueous sub-surface life, or (c) biological processes from surface life thriving in the hydrocarbon lakes.

---

[1] though there is a staggering diversity of Earth environments.

This is why the liquid water habitable zone, for all its limitations, is presently such a powerful, universal criterion. Our detailed knowledge of life on Earth provides the high-resolution context needed to evaluate life elsewhere. If we want to look for life outside the liquid water habitable zone, then both the possibility of this exotic life, and the features it produces, must be established first within the solar system. The search for the exotic, must start close to home.

**FIGURES**

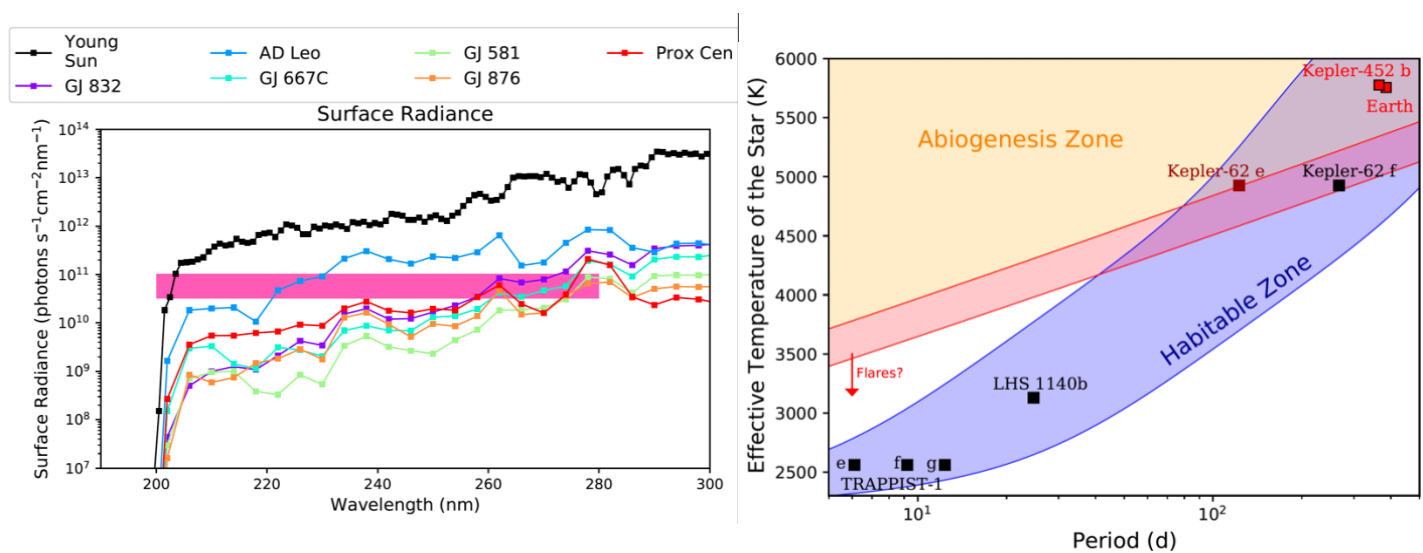

**Figure 1: (Left)** The surface irradiance as a function of wavelength, as experienced at the surface of a rocky planets with a 1 bar CO2/N2 atmosphere, within the liquid water habitable zone of its host star. All the stars represented, besides the Young Sun, are M Dwarfs, with effective temperatures <3500 K. A pink rectangle represents the flux that delimits the abiogenesis zone (Rimmer et al. 2018). Figure modified from Figure 2 of Ranjan et al. (2017), which is © AAS and Reproduced with permission. The abiogenesis zone (orange region), with error bars representing experimental uncertainty (red region), and liquid water habitable zone (blue region), as a function of the effective temperature of the star and the orbital period of a planet around that star, based on Fig. 4 of Rimmer et al. (2018), reproduced under Creative Commons. The surface UV calculations of Ranjan et al. (2017) indicate that low-mass or inactive M-dwarfs may lack the steady-state UV required to power the cyanosulfidic chemistry, while Rimmer et al. (2018) indicate that all known M-dwarfs fall short of this constraint. M-dwarf planets may offer the exciting possibility of testing whether UV light in general and the cyanosulfidic chemistry in particular is required for the origin of life, but confirming this possibility and identifying which M-dwarf planets, if any, could fulfill this role, will require further theoretical, experimental, and observational work.

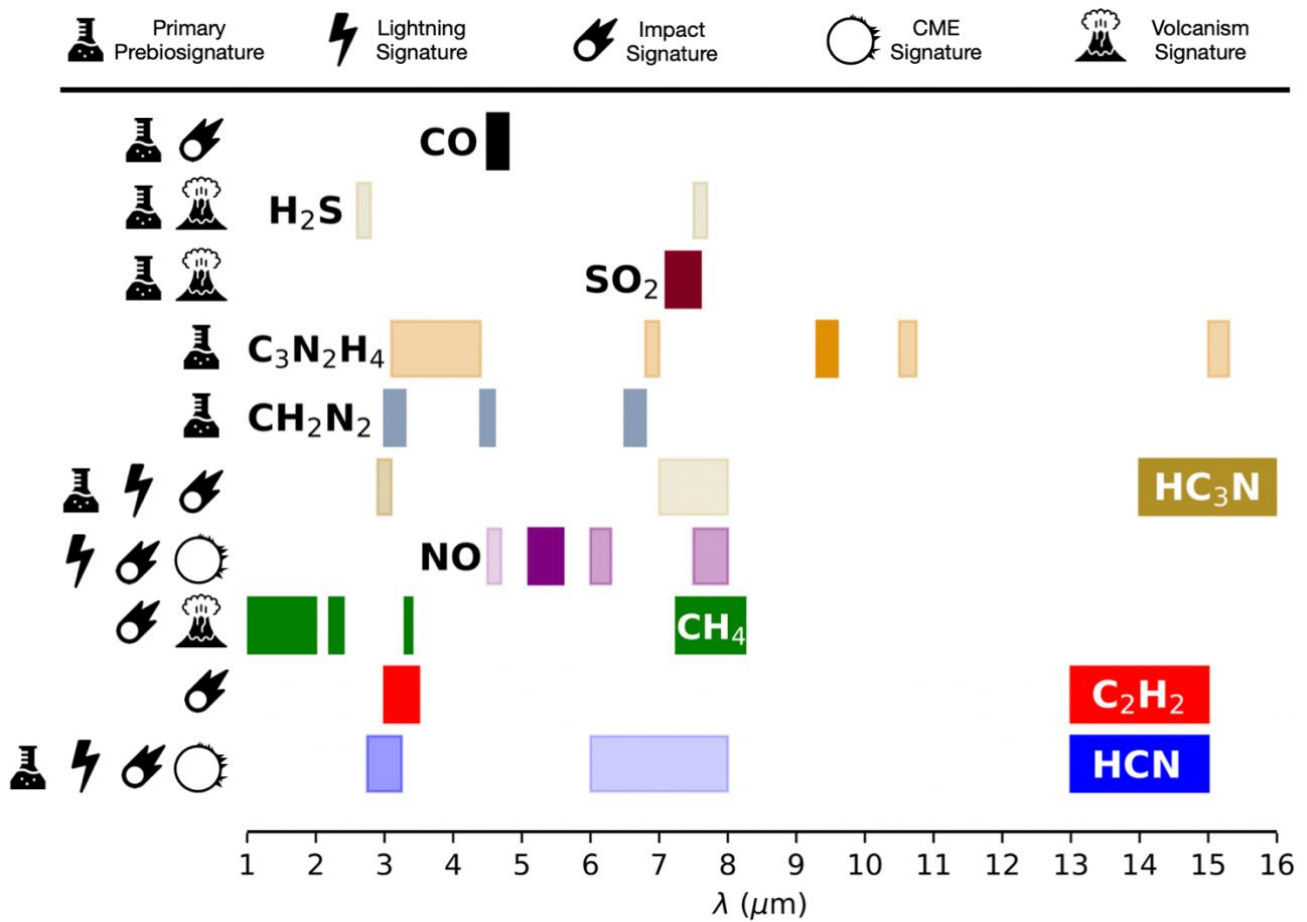

**Figure 2:** Spectral features for prebiosignatures as a function of wavelength. The species include carbon monoxide (CO), hydrogen sulfide ($H_2S$), sulfur dioxide ($SO_2$), imidazole ($C_3N_2H_4$), cyanamide ($CH_2N_2$), cyanoacetylene ($HC_3N$), nitric oxide (NO), methane ($CH_4$), acetylene ($C_2H_2$) and hydrogen cyanide (HCN). Icons represent whether the signatures are primary prebiosignatures (species directly relevant for prebiotic chemistry), secondary prebiosignatures (species that are produced by events that are relevant for prebiotic chemistry, such as volcanism, lightning, impacts or particles ejected from the star) or both. The features are strong to weak in proportion to the transparency of the rectangles. Spectra and relevant references can be found at https://doi.org/10.7910/DVN/UTFCHI.

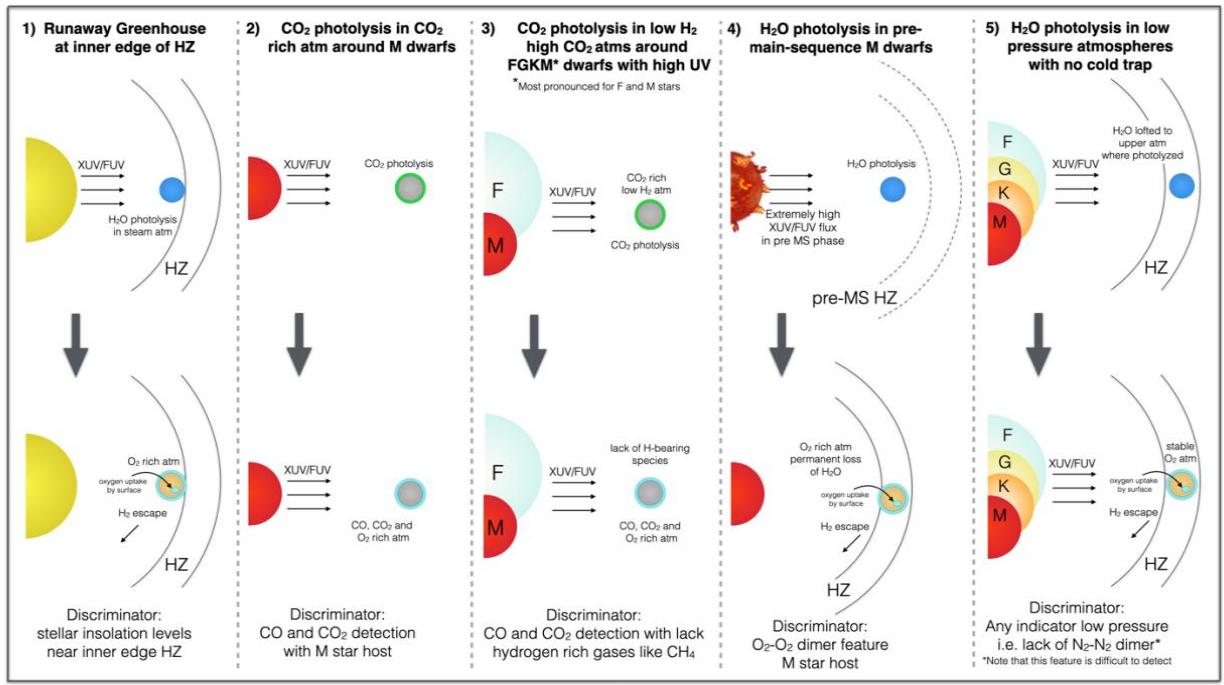

**Figure 3:** Molecular oxygen (O₂) as a biosignature, false positives and discriminants. The first panel shows how a runaway greenhouse can produce significant amounts of O₂. The second panel illustrates how O₂ can be produced on rocky planets around M dwarfs by photodissociation. The third panel shows how O₂ can be produced in H₂-poor CO₂-rich atmospheres of rocky planets around FGKM dwarfs. The fourth panel shows how O₂ can be produced from the FUV/XUV fluxes of pre-main-sequence M dwarfs. And the fifth panel shows how O₂ can be produced on planets with low-pressure atmospheres and without cold traps. Discriminators are given at the bottom of each panel.